\documentclass[]{book}

\usepackage[utf8]{inputenc}
\usepackage{graphicx}
\usepackage[numbers]{natbib}
\usepackage{wileySTM}
\usepackage[\ifnum\pdfoutput=1breaklinks\fi]{hyperref}
\usepackage{braket}
\usepackage{tikz}
\usetikzlibrary{shapes.geometric, arrows}
\title{}
\author{}

\graphicspath{{figures/}}

\tikzstyle{startstop} = [rectangle, rounded corners, minimum width=3cm, minimum height=1cm,text centered, draw=black, fill=blue!30]
\tikzstyle{process}   = [rectangle, minimum width=3cm, minimum height=1cm, text centered, draw=black, fill=orange!30]
\tikzstyle{decision}  = [diamond,   minimum width=3cm, minimum height=1cm, text centered, draw=black, fill=green!30]
\tikzstyle{arrow}     = [thick,->,>=stealth]

\begin{document}

\chapterauthor{Sebastian Wouters, Carlos A. Jim\'enez-Hoyos and Garnet K.-L. Chan}
\chapter{Five years of density matrix embedding theory}

Density matrix embedding theory (DMET) describes finite fragments in the presence of a surrounding environment.
In contrast to most embedding methods, DMET explicitly allows for quantum entanglement between both.
In this chapter, we discuss both the ground-state and response theory formulations of DMET, and review several applications. In addition, a proof is given that the local density of states can be obtained by working with a Fock space of bath orbitals.

\section{Quantum entanglement} \label{entanglement_introduction}
In this section, we review nomenclature and several concepts from quantum information theory, which are necessary to follow the discussion on density matrix embedding theory. Quantum many-body theory can be formulated in a Fock space of single-particle states. In quantum chemistry, the single-particle states are orbitals, each with $d$ possible occupations. For spin-orbitals there are $d = 2$ possible occupations: empty or occupied. For spatial orbitals there are $d=4$ possible occupations: empty, occupied with one electron with spin-projection $s^z = \pm \frac{1}{2}$, or occupied with two electrons.

Consider a bipartition of the total Hilbert space (the Fock space of all orbitals) into two subsystems A and B (the Fock spaces of orbital groups A and B). One can think, for example, about the left and right halves of a polyene, or the split-up into active and external orbital spaces in a complete active space (CAS) calculation. Let $L_{\text{A}}$ ($L_{\text{B}}$) denote the number of orbitals in subsystem A (B). The Hilbert space of subsystem A (B) has size $N_{\text{A}} = d^{L_{\text{A}}}$ ($N_{\text{B}} = d^{L_{\text{B}}}$). Let $\{ \ket{i}_{\text{A}} \}$ $\left(\{ \ket{j}_{\text{B}} \}\right)$ denote a particular many-body basis of size $N_{\text{A}}$ ($N_{\text{B}}$) for subsystem A (B). One can think, for example, about all possible Slater determinants formed by considering all $d$ possible occupations of all $L_{\text{A}}$ ($L_{\text{B}}$) orbitals in subsystem A (B). Any unitary rotation of this many-body basis is of course as good a choice.

The total Hilbert space $\mathcal{H}$ is the tensor product of the Hilbert spaces of subsystems A and B: $\mathcal{H} = \mathcal{H}_{\text{A}} \otimes \mathcal{H}_{\text{B}}$. $\mathcal{H}$ is spanned by $d^{L_{\text{A}} + L_{\text{B}}} = N_{\text{A}}N_{\text{B}}$ many-body basis states. One particular choice is $\{ \ket{i}_{\text{A}} \ket{j}_{\text{B}} \}$. Any state in $\mathcal{H}$ can be expressed as
\begin{equation}
 \ket{\Psi} = \sum\limits_{i }^{N_{\text{A}}} \sum\limits_{j}^{N_{\text{B}}} C_{ij} \ket{i}_{\text{A}} \ket{j}_{\text{B}}.
\end{equation}
By considering the singular value decomposition of the coefficient tensor
\begin{equation}
C_{ij} = \sum\limits_{\alpha}^{\text{min}(N_{\text{A}},N_{\text{B}})} U_{i\alpha} \lambda_{\alpha} V^{\dagger}_{\alpha j},
\end{equation}
this state can be rewritten as
\begin{equation}
 \ket{\Psi} = \sum\limits_{i}^{N_{\text{A}}} \sum\limits_{j}^{N_{\text{B}}} \sum\limits_{\alpha}^{\text{min}(N_{\text{A}},N_{\text{B}})} U_{i\alpha} \lambda_{\alpha} V^{\dagger}_{\alpha j} \ket{i}_{\text{A}} \ket{j}_{\text{B}} = \sum\limits_{\alpha}^{\text{min}(N_{\text{A}},N_{\text{B}})} \lambda_{\alpha} \ket{\alpha}_{\text{A}} \ket{\alpha}_{\text{B}}. \label{schmidt_decomp}
\end{equation}
The unitary transformation $U_{i\alpha}$ $\left(V^*_{j\beta} = V^{\dagger}_{\beta j}\right)$ rotates the many-body basis $\{ \ket{i}_{\text{A}} \}$ $\left(\{ \ket{j}_{\text{B}} \}\right)$ to the new many-body basis $\{ \ket{\alpha}_{\text{A}} \}$ $\left(\{ \ket{\beta}_{\text{B}} \}\right)$. The particular form of $\ket{\Psi}$ in Eq.~\eqref{schmidt_decomp} is called the Schmidt decomposition. When $\mathcal{H}_{\text{A}}$ and $\mathcal{H}_{\text{B}}$ are of different sizes, the Schmidt decomposition allows for a compact representation of $\ket{\Psi}$. When, for example, $N_{\text{A}} < N_{\text{B}}$ only $N_A$ many-body basis states in $\mathcal{H}_{\text{B}}$ are needed to represent $\ket{\Psi}$ exactly.

When only one of the Schmidt or singular values $\lambda_{\alpha}$ is nonzero, the state $\ket{\Psi}$ is factorizable with respect to the given bipartition, and the two subsystems are said to be unentangled. When several of the Schmidt values are nonzero, the state $\ket{\Psi}$ is not factorizable with respect to the given bipartition, and the two subsystems are said to be entangled.

In CAS wavefunctions, the occupations of the external orbitals are invariant over all determinants with nonzero coefficients. Such a wavefunction is factorizable with respect to the bipartition into active and external orbital spaces, and these two subsystems are therefore unentangled. In a polyene, the $\pi$-conjugation requires a multireference description. The occupations of the orbitals in the left (or right) half of the polyene differ over the determinants with nonzero coefficients. The multireference ground-state wavefunction of a polyene is therefore not factorizable with respect to the bipartition into left and right halves, and these subsystems are therefore entangled.

\section{Density matrix embedding theory} \label{dmet_general_introduction}
Suppose one is interested in the properties of a small subsystem of the whole quantum system. Synonyms for this subsystem are the \textit{impurity}, \textit{cluster}, or \textit{fragment}. While the impurity is most often a specific region in space, it can be any subsystem of the total Hilbert space. Other examples are a band in momentum space, a particular non-localized set of orbitals in a molecule, or even an incomplete many-body basis which cannot be rewritten as a Fock space of orbitals. In this chapter, we will limit ourselves to a Fock space of impurity orbitals. The region outside of the impurity is called the \textit{environment}.

The general goal in embedding theories is to obtain the properties of interest of the impurity, without doing expensive calculations on the whole quantum system. (Part of) the environment can for example be replaced with a solvent model, static charges, or a mean-field description which is unentangled with the impurity. For many systems such a description gives accurate results. However, when there is static correlation between the impurity and the environment, these methods will fail.

In DMET, the wavefunction for the impurity plus environment is written as in Eq.~\eqref{schmidt_decomp}. If the exact wavefunction $\ket{\Psi}$ is known, the decomposition from the previous section can be followed to construct a Schmidt basis for the environment. But this requires \textit{a priori} knowledge of $\ket{\Psi}$, and then we wouldn't have to use an embedding theory in the first place. The idea of DMET is to embed the impurity A in an approximate bath B. The bath can be thought of as $N_{\text{A}}$ approximate many-body basis states for the environment. Solving the impurity plus bath system is called the \textit{embedded} problem. Various possibilities exist to obtain a bath space. In this chapter, we will limit ourselves to a Fock space of bath orbitals, obtained from a \textit{low-level} particle-number conserving mean-field wavefunction \cite{geraldPRL, geraldJCTC}. This is however not the only possibility. One can also work with single-particle states from Hartree-Fock-Bogoliubov theory \cite{boxiaoHubbard, simons_collaboration} or with many-body basis states from a Schmidt decomposition of anti-symmetrized geminal power (AGP) wavefunctions for electrons \cite{troyGeminals}, coherent state wavefunctions for phonons \cite{barbara}, or block-product states for spins \cite{spinsystemPRB}. Although there are real benefits to using the most accurate feasible wavefunctions for the bath construction, it is also convenient to recycle the large number of existing quantum many-body solvers when solving the embedded problem. They rely on Fock spaces of orbitals, as they are constructed for second-quantized Hamiltonians.

Once initial bath states have been obtained from a so-called \textit{low-level} wavefunction, they can be left unoptimized. We call this \textit{single-shot} DMET embedding. Alternatively, once an Ansatz for the bath states is chosen, its parameters can be variationally optimized with a \textit{high-level} method \cite{spinsystemPRB}. Another option, which is most often used when the entire system is tiled with impurities, see Fig. \ref{system_tiling}, is to introduce a DMET correlation potential $\hat{u}$ to link a single low-level wavefunction, which produces the bath states for the different impurities, with the high-level wavefunctions of these impurities \cite{geraldPRL, geraldJCTC}. The Hamiltonian which yields the low-level wavefunction is augmented with $\hat{u}$, and $\hat{u}$ itself is optimized in order to match properties in the low- and high-level wavefunctions. In DMET, as the name indicates, (parts of) the density matrices of the low- and high-level wavefunctions are matched. In this chapter, we will limit ourselves to the latter strategy.

Dynamical mean-field theory (DMFT) \cite{PhysRevLett.62.324, PhysRevLett.69.1240, RevModPhys.68.13, zgidDMFT} is another successful method to embed an impurity in an environment. Instead of setting up a self-consistent scheme to match density matrices, the Green's function of the impurity is determined self-consistently by fitting the frequency-dependent hybridization function. The fit parameters turn up as bath orbitals which are coupled to the impurity. The result of DMFT is a correlated frequency-dependent Green's function, which should still be integrated over a contour to obtain ground-state energies. For ground-state properties, DMET provides a computationally simpler and cheaper alternative with similar accuracy. But DMET is not at all limited to ground-state properties, as will be discussed in Sect.~\ref{section_response_theory}.

\section{Bath orbitals from a Slater determinant} \label{bath_orbital_section}

Consider a Slater determinant approximation $\ket{\Phi_0}$ for the ground-state of the full system. In second quantization, it can be written as
\begin{equation}
\ket{\Phi_0} = \prod\limits_{\mu}^{N_{\text{occ}}} \hat{a}^{\dagger}_{\mu} \ket{-}. \label{HFwfn}
\end{equation}
The $N_{\text{occ}}$ occupied spin-orbitals are denoted by $\mu\nu$, the $L$ orthonormal spin-orbitals for the impurity and its environment by $klmn$, and the orthonormal impurity and bath orbitals by $pqrs$. There are $L_{\text{A}}$ orbitals in the impurity A. In what follows, we initially assume that $N_{\text{occ}} \geq L_{\text{A}}$.

The occupied orbitals can be written in terms of the impurity and environment orbitals:
\begin{equation}
\hat{a}^{\dagger}_{\mu} = \sum\limits_{k}^{L} \hat{a}_k^{\dagger} C_{k\mu}.
\end{equation}
The physical wavefunction represented by Eq.~\eqref{HFwfn} does not change when the occupied orbitals are internally rotated. Ref.~\citenum{geraldJCTC} discusses how this freedom can be used to split the occupied orbital space into two parts: orbitals with and without overlap on the impurity. This construction can be understood by means of a singular value decomposition. Consider the occupied orbital coefficient block with indices on the impurity: $k \in {\text{A}}$. The singular value decomposition of the $L_{\text{A}} \times N_{\text{occ}}$ coefficient block $C_{k\mu}$ yields an occupied orbital rotation matrix $V_{\mu p}$:
\begin{equation}
C_{k\mu}(k \in {\text{A}}) = \sum\limits_{p}^{L_{\text{A}}} U_{kp} \lambda_{p} V^{\dagger}_{p\mu},
\end{equation}
which can be made square by adding $N_{\text{occ}}-L_{\text{A}}$ extra columns: $W = \left[ V \widetilde{V} \right]$. The occupied orbital space can now be rotated with the $N_{\text{occ}} \times N_{\text{occ}}$ matrix $W$:
\begin{equation}
\hat{a}^{\dagger}_{p} = \sum\limits_{\mu}^{N_{\text{occ}}} \hat{a}^{\dagger}_{\mu} W_{\mu p} = \sum\limits_{\mu}^{N_{\text{occ}}} \sum\limits_{k}^{L} \hat{a}_k^{\dagger} C_{k\mu} W_{\mu p} = \sum\limits_{k}^{L} \hat{a}_k^{\dagger} \widetilde{C}_{kp}.\label{bathorbsconstructattempt1}
\end{equation}
Of the rotated occupied orbitals, only $L_{\text{A}}$ have nonzero overlap with the impurity:
\begin{equation}
\widetilde{C}_{kp}(k \in {\text{A}}) = \sum\limits_{\mu}^{N_{\text{occ}}} \sum\limits_{q}^{L_{\text{A}}} U_{kq} \lambda_{q}  V^{\dagger}_{q\mu} W_{\mu p} =
\begin{cases} 
   U_{kp} \lambda_{p} & \text{if } p \leq L_{\text{A}} \\
   0       & \text{otherwise}
  \end{cases}. \label{rotated_mo_orbitals}
\end{equation}
The $N_{\text{A}}$ bath states $\left\{ \ket{\alpha}_{\text{B}} \right\}$ with nonzero Schmidt number can be found by diagonalizing
\begin{equation}
\hat{\rho}_{\text{B}} = \sum\limits_{i}^{N_{\text{A}}} \braket{ i \mid_{\text{A}} \Phi_0} \braket{ \Phi_0 \mid i}_{\text{A}}.
\end{equation}
The occupied orbitals without weight on the impurity will always be fully occupied for all projections $\braket{ \Phi_0 \mid i}_{\text{A}}$. In contrast, the occupied orbitals with weight on the impurity can become partially occupied. The DMET bath construction then yields a CAS Ansatz:
\begin{enumerate}
 \item The external core orbitals are the $N_{\text{occ}} - L_{\text{A}}$ occupied orbitals without overlap on the impurity.
 \item The active orbitals consist of the impurity and bath orbitals. The bath orbitals are the $L_{\text{A}}$ occupied orbitals with overlap on the impurity, after projection onto the environment:
 \begin{equation}
 \hat{a}^{\dagger}_{r} (r \leq L_{\text{A}}) = \sum\limits_{k > L_{\text{A}}} \hat{a}_{k}^{\dagger} \frac{ \widetilde{C}_{k r} }{ \sqrt{\sum\limits_{l > L_{\text{A}}} \mid \widetilde{C}_{l r} \mid^2 } } = \sum\limits_{k > L_{\text{A}}} \sum\limits_{\mu}^{N_{\text{occ}}} \hat{a}_{k}^{\dagger} \frac{ C_{k \mu} V_{\mu r} }{ \sqrt{ 1 - \lambda_r^2 } }. \label{bathorbitalconstruction}
 \end{equation}
 \item The external virtual orbitals are the $L - N_{\text{occ}} - L_A$ unoccupied orbitals, after projection onto the environment.
\end{enumerate}
Note that for this CAS Ansatz, the impurity and the environment can be entangled, as the latter is represented by $L_{\text{A}}$ bath orbitals in the active space. If the active space comprises solely of the impurity orbitals, this is not the case.

To find the bath orbitals, the overlap of the occupied orbitals with the impurity is diagonalized in Ref.~\citenum{geraldJCTC}:
\begin{equation}
S_{\mu\nu} = \sum\limits_{k}^{L_{\text{A}}} C^{\dagger}_{\mu k}C_{k\nu} = \sum\limits_{p}^{L_{\text{A}}} V_{\mu p} \lambda_{p}^2 V^{\dagger}_{p \nu}. \label{origOverlap}
\end{equation}
At most $L_{\text{A}}$ eigenvalues of $S_{\mu\nu}$ are nonzero. The $L_{\text{A}}$ corresponding eigenvectors yield the bath orbitals in Eq.~\eqref{bathorbitalconstruction}. A completely equivalent construction of the bath orbitals is to consider the environment block of the mean-field density matrix:
\begin{equation}
D_{kl} (kl > L_{\text{A}}) = \braket{\Phi_0 \mid \hat{a}_l^{\dagger} \hat{a}_k \mid \Phi_0} = \sum\limits_{\mu}^{N_{\text{occ}}} C_{k \mu} C^{\dagger}_{\mu l} = \sum\limits_{p}^{N_{\text{occ}}} \widetilde{C}_{k p} \widetilde{C}^{\dagger}_{p l}.
\end{equation}
MacDonald's theorem \cite{PhysRev.43.830} states that when a row and the corresponding column of a Hermitian matrix are removed, the eigenvalues of the new matrix lie in between the original eigenvalues. Therefore at most $L_{\text{A}}$ eigenvalues of the $(L - L_{\text{A}}) \times (L - L_{\text{A}})$ environment subblock $D_{kl} (kl > L_{\text{A}})$ will lie in between 0 and 1. The corresponding eigenvectors are the orthonormal bath orbitals from Eq.~\eqref{bathorbitalconstruction}. The $N_{\text{occ}} - L_{\text{A}}$ eigenvectors with eigenvalue 1 are the external core orbitals. The $L - N_{\text{occ}} - L_{\text{A}}$ eigenvectors with eigenvalue 0 are the external virtual orbitals.

The overlap matrix $S_{\mu\nu}$ in Eq.~\eqref{origOverlap} is a projector of the occupied orbitals onto the impurity. Analogously, $D_{kl}(kl > L_{\text{A}})$ is a projector of the environment orbitals onto the occupied orbitals. Eigenvectors with partial weight signal occupied orbitals with support on both the impurity and the environment, \textit{i.e.} they are entangled orbitals.

We have assumed that $N_{\text{occ}} \geq L_{\text{A}}$. However, for very large basis sets this will not be the case. Because bath orbitals arise from the decomposition of occupied orbitals with support on both the impurity and the environment, there can be at most $N_{\text{occ}}$ bath orbitals. Even if $N_{\text{occ}} \gtrsim L_{\text{A}}$, the DMET bath orbital selection will try to add low-lying core electrons corresponding to atoms in the environment into the bath space, an undesired effect. One way to circumvent this problem is to define for the impurity a core, valence, and virtual orbital space (in the molecular sense) and to try to find bath orbitals only for the valence orbitals. In Ref. \citenum{seb_dmet} we show that Knizia's intrinsic atomic orbitals \cite{iao_gerald2} are particularly suited for this strategy. Another way would be to take the most correlated orbitals, typically one bath orbital per chemical bond. This boils down to Sun's optimal QM/MM boundary scheme \cite{qimingJCTC}. In the remainder we will use $L_{\text{B}}$ to denote the number of bath orbitals. All other environment orbitals are restricted to be fully occupied or empty. The deficit in electron number between the external core orbitals and $N_{\text{occ}}$ is the number of electrons in the active space $N_{\text{act}}$.

\section{The embedding Hamiltonian} \label{embedded_hamiltonian}

\begin{figure}
\centering
 \includegraphics[width=0.5\textwidth]{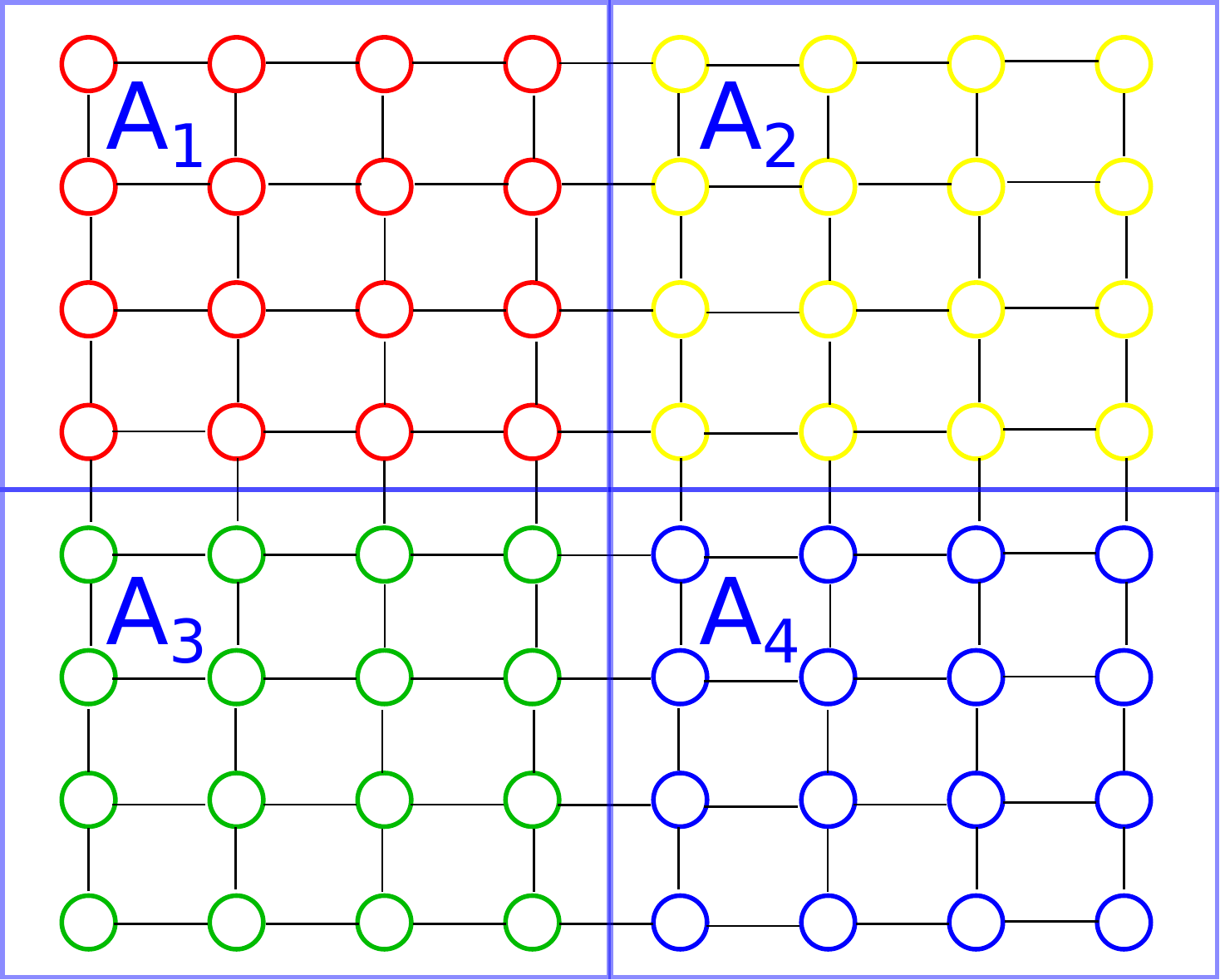}
 \caption{\label{system_tiling} The total system is tiled with impurities. When the circles denote orbitals, each impurity is a Fock space of impurity orbitals.}
\end{figure}

The Hamiltonian for the total system can be written as:
\begin{equation}
\hat{H} = E_{\text{nuc}} + \sum\limits_{kl}^L t_{kl} \hat{a}_k^{\dagger} \hat{a}_l + \frac{1}{2} \sum\limits_{klmn}^L (kl|mn) \hat{a}_k^{\dagger} \hat{a}_m^{\dagger}  \hat{a}_n \hat{a}_l, \label{orig_ham_eq}
\end{equation}
where $t_{kl}$ and $(kl|mn)$ are diagonal in the spin indices of spin-orbitals $k$ and $l$, and $(kl|mn)$ is diagonal in the spin indices of $m$ and $n$ as well. We assume fourfold permutation symmetry $(kl|mn) = (mn|kl) = (lk|nm)$ for the electron repulsion integrals.

Suppose the total system is tiled with impurities, see Fig. \ref{system_tiling}. For each impurity A$_x$ a Hermitian one-particle operator $\hat{u}^x$ is introduced which acts solely within the impurity A$_x$:
\begin{equation}
 \hat{u}^x = \sum\limits_{kl}^{L_{\text{A}_x}} u^x_{kl} \hat{a}_k^{\dagger} \hat{a}_l.
\end{equation}
The sum of all these one-particle operators forms the DMET correlation potential:
\begin{equation}
 \hat{u} = \sum\limits_{x} \hat{u}^x.
\end{equation}
With $\vec{u}$ we denote all of its independent variables. The mean-field low-level wavefunction $\ket{\Phi_0(\vec{u})}$ is obtained as either the mean-field solution of $\hat{H} + \hat{u}$, or as the eigenfunction of a one-particle operator $\hat{h} + \hat{u}$. For local electron repulsion integrals, for example in the Hubbard model, the one-particle operator $\hat{h}$ is the hopping matrix. For nonlocal quartic electron repulsion integrals, which arise in quantum chemistry, the one-particle operator $\hat{h}$ is the Fock operator corresponding to $\hat{H}$.

All variables $\vec{u}$ influence the bath orbitals of each impurity A$_x$. Together with the impurity orbitals of A$_x$, these bath orbitals form an active space. The $N_{\text{act}}^{x}$ active space electrons interact with the external core electrons as well, and therefore it is important to take them into account via their density matrix:
\begin{equation}
 D_{kl}^{\text{core},x} = \sum\limits_{p \in \text{core}} \widetilde{C}_{k p} \widetilde{C}^{\dagger}_{p l}.
\end{equation}
The one-electron integrals of the embedding Hamiltonian
\begin{equation}
 \hat{H}^{\text{emb},x} = \sum\limits_{pq}^{L_{\text{A}_x} + L_{\text{B}_x}} h^x_{pq} \hat{a}^{\dagger}_p \hat{a}_q + \frac{1}{2} \sum\limits_{pqrs}^{L_{\text{A}_x} + L_{\text{B}_x}} (pq|rs) \hat{a}^{\dagger}_p \hat{a}^{\dagger}_r \hat{a}_s \hat{a}_q - \mu_{\text{glob}}  \sum\limits_{r}^{L_{\text{A}_x}}  \hat{a}_r^{\dagger} \hat{a}_r
\end{equation}
contain the Coulomb and exchange contributions due to the external core electrons:
\begin{equation}
 h^x_{kl} = t_{kl} + \sum\limits_{mn}^{L} \left[ (kl|mn) - (kn|ml) \right] D_{mn}^{\text{core},x} \xrightarrow{ \widetilde{C}_{kp} } h^x_{pq}.
\end{equation}
This embedding Hamiltonian has an \textit{interacting bath}, as there are electron repulsion integrals for the bath. The DMET correlation potential appears only indirectly through its effect on the form of the bath and external core orbitals. To ensure that the total number of electrons in all impurities A$_x$ adds up to $N_{\text{occ}}$, it becomes necessary to introduce a global chemical potential $\mu_{\text{glob}}$ for the impurity orbitals. This global chemical potential is independent of the specific impurity A$_x$.

For local electron repulsion integrals, for example in the Hubbard model, the core electrons don't have Coulomb or exchange interactions with the impurity orbitals. In this case, a simpler version of DMET is commonly used, where we replace the electron repulsion in the bath, due to the external core and bath electrons, by the DMET correlation potential rotated to the bath orbitals:
\begin{equation}
 \hat{H}^{\text{emb},x} = \sum\limits_{pq}^{L_{\text{A}_x} + L_{\text{B}_x}} t_{pq} \hat{a}^{\dagger}_p \hat{a}_q + \sum\limits_{pq}^{L_{\text{B}_x}} u_{pq} \hat{a}^{\dagger}_p \hat{a}_q + \frac{1}{2} \sum\limits_{pqrs}^{L_{\text{A}_x}} (pq|rs) \hat{a}^{\dagger}_p \hat{a}^{\dagger}_r \hat{a}_s \hat{a}_q. \label{hubb_nibath}
\end{equation}
On the impurity, the original electron repulsion integrals act. This embedding Hamiltonian has a \textit{noninteracting bath}, as there are no electron repulsion integrals for the bath. The DMET correlation potential then has a triple role: it determines the form of the bath and external core orbitals, it represents the electron repulsion in the bath, and it ensures that the total number of electrons in all impurities A$_x$ adds up to $N_{\text{occ}}$.

The ground state $\ket{\Psi_x}$ of the embedding Hamiltonian $\hat{H}^{\text{emb},x}$ for impurity A$_x$ is calculated with a high-level method, typically full configuration interaction \cite{geraldPRL, geraldJCTC}, the density-matrix renormalization group \cite{boxiaoHubbard, qiaoniPRB}, or coupled-cluster theory \cite{bulikJCP}. In order to calculate expectation values of interest, it should be possible to obtain one- and two-particle density matrices of the active space with the high-level method:
\begin{eqnarray}
D_{rs}^x & = & \braket{ \Psi_x \mid \hat{a}^{\dagger}_r \hat{a}_s \mid \Psi_x }, \\
P^x_{pq|rs} & = & \braket{ \Psi_x \mid \hat{a}^{\dagger}_p \hat{a}^{\dagger}_r \hat{a}_s \hat{a}_q \mid \Psi_x }.
\end{eqnarray}
For \textit{local} operators, when all orbital indices correspond to one particular impurity, the density matrices of that impurity are used to obtain the expectation values. For \textit{nonlocal} operators, a democratic partitioning of the Hermitian expectation values is used. Suppose, for example, that orbital $i$ ($j$) is part of impurity A$_x$ (A$_y$):
\begin{equation}
\braket{ \hat{a}_i^{\dagger} \hat{a}_j +  \hat{a}_j^{\dagger} \hat{a}_i } = 
\braket{ \Psi_{x} | \hat{a}_i^{\dagger} \hat{a}_j | \Psi_{x} } + 
\braket{ \Psi_{y} | \hat{a}_j^{\dagger} \hat{a}_i | \Psi_{y} }.
\end{equation}
By convention, the density matrix of the impurity corresponding to the first index is used. For an embedding Hamiltonian with interacting bath, this gives rise to the following formula for the total energy:
\begin{align}
E_{\text{tot}} & =  E_{\text{nuc}} + \sum\limits_x E_{x}, \label{energyfragment_pre}\\
E_{x} & =  \sum\limits_{p}^{L_{{\text{A}}_x}} \left( \sum\limits_{q}^{L_{{\text{A}}_x}+L_{{\text{B}}_x}} \frac{t_{pq} + h_{pq}^x}{2} D_{pq}^{x} + \frac{1}{2} \sum\limits_{qrs}^{L_{{\text{A}}_x}+L_{{\text{B}}_x}} (pq|rs) P_{pq|rs}^{x} \right). \label{energyfragment}
\end{align}
The one-electron integrals in Eq.~\eqref{energyfragment} avoid the double counting of Coulomb and exchange contributions of the external core electrons when they arise in the active spaces of other impurities. The factor $\frac{1}{2}$ is similar to the difference between the Fock operator and energy expressions in HF theory.

Note that DMET energies are not variational, because density matrices of different high-level calculations, each in their own active space consisting of impurity and bath orbitals, enter in Eqs.~\eqref{energyfragment_pre}-\eqref{energyfragment}.

\section{Self-consistency} \label{section_self_consistency}
The DMET correlation potential $\hat{u}$ is determined by matching (parts of) the low-level and high-level one-particle density matrices as closely as possible. These parts can be: for each impurity the full one-particle density matrix in its active space; for each impurity the impurity block of the one-particle density matrix; for each impurity the diagonal of the impurity block of the one-particle density matrix; or just the total number of electrons in all impurities. In order to have a well-posed optimization problem, the number of variables in $\vec{u}$ should not exceed the number of independent expectation values which are matched. For the four cases discussed above, respectively the full DMET correlation potential; the full DMET correlation potential; the diagonal of the DMET correlation potential; or just the global chemical potential $\mu_{\text{glob}}$ are optimized. The DMET correlation potential $\hat{u}$ and the global chemical potential $\mu_{\text{glob}}$ can be obtained by least-squares minimization of:
\begin{eqnarray}
 \Delta^x_{pq}(\vec{u}) & = & D^{\text{low},x}_{pq}(\vec{u}) - D^{\text{high},x}_{pq}, \\
 \Delta_N(\mu_{\text{glob}}) & = & N_{\text{tot}}(\mu_{\text{glob}}) - N_{\text{occ}}.
\end{eqnarray}
Note that the high-level density matrix $D^{\text{high},x}_{pq}$ also depends on $\vec{u}$, either solely through the form of the bath and external core orbitals for an interacting bath, or also directly through the Hamiltonian matrix elements in Eq.~\eqref{hubb_nibath} for a noninteracting bath. However, thus far the DMET correlation potential has always been optimized while keeping the high-level density matrix $D^{\text{high},x}_{pq}$ fixed. The appendix in Ref. \citenum{seb_dmet} discusses how to obtain analytic gradients of $D^{\text{low},x}_{pq}(\vec{u})$ with respect to $\vec{u}$.

As discussed extensively in Ref.~\citenum{troyGeminals}, trying to match (a part of) a mean-field density matrix with (a part of) a high-level correlated density matrix is not always possible because the former is idempotent while the latter does not have to be. Thus, obtaining a correlation potential for which the cost function becomes zero is therefore not always possible.

$\vec{u}$ can be optimized per impurity A$_x$ by projecting the mean-field problem into its active space. Only the elements $u^x_{kl}$ corresponding to the specific impurity A$_x$ are then optimized. The low-level wavefunction should then be calculated in a small orbital space, but as all elements $\vec{u}$ can influence all bath spaces, this approach is prone to limit cycles and slow convergence due to overshooting. It might therefore be better to first solve the mean-field problem in the total system, and to subsequently project its density matrix to the different active spaces \cite{seb_dmet}. The desired parts of the density matrices are then matched simultaneously for all impurities. Stationary points of the latter approach will also be stationary points of the former.

In Ref. \citenum{seb_dmet} we discuss how the function minimization of the difference of the one-particle density matrices with respect to $\hat{u}$ can be recast into a functional optimization problem with respect to $\ket{\Phi_0}$.

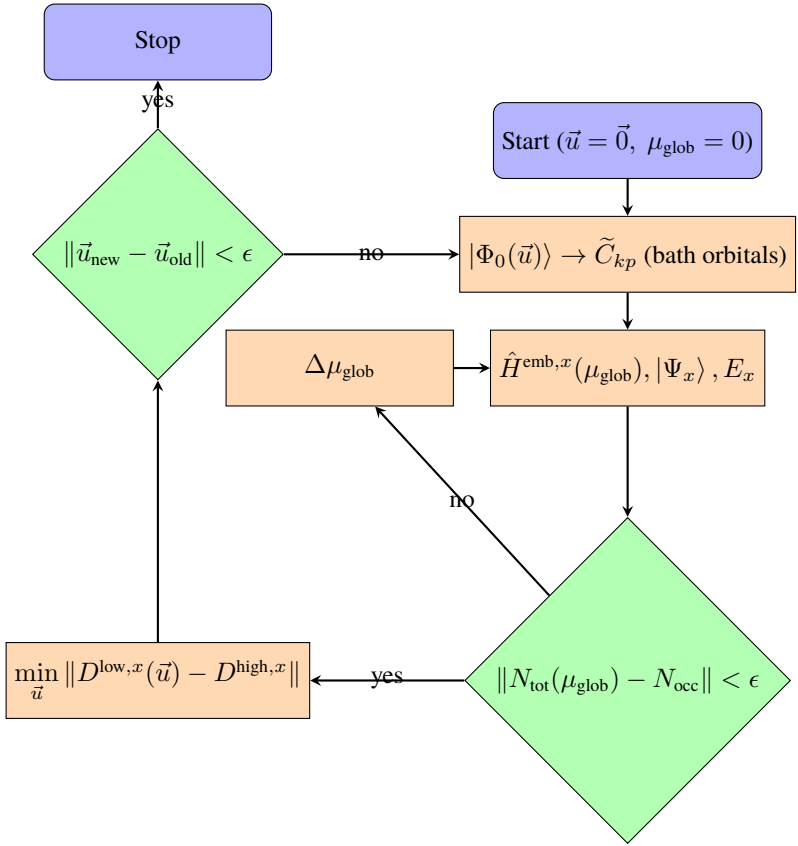
\begin{figure}
\begin{tikzpicture}[node distance=2cm]
\node (start) [startstop] {Start ($\vec{u}=\vec{0},~ \mu_{\text{glob}}=0$)};
\node (pro1) [process, below of=start, yshift=0.5cm] {$\ket{\Phi_0(\vec{u})} \rightarrow \widetilde{C}_{kp}$ (bath orbitals)};
\node (pro2) [process, below of=pro1, yshift=0.5cm] {$\hat{H}^{\text{emb},x}(\mu_{\text{glob}}), \ket{\Psi_x}, E_x $};
\node (dec1) [decision, below of=pro2, yshift=-2.12cm] {$ \| N_{\text{tot}}(\mu_{\text{glob}}) - N_{\text{occ}} \| < \epsilon $};
\node (pro3a) [process, left of=pro2, xshift=-1.8cm] {$\Delta \mu_{\text{glob}}$};
\node (pro3b) [process, left of=dec1, xshift=-4.2cm] {$\min\limits_{\vec{u}} \|D^{\text{low},x}(\vec{u}) - D^{\text{high},x}\|$};
\node (dec4) [decision, left of=pro1, xshift=-4.2cm] {$ \| \vec{u}_{\text{new}} - \vec{u}_{\text{old}} \| < \epsilon$};
\node (stop) [startstop, above of=dec4, yshift=0.8cm] {Stop};

\draw [arrow] (start) -- (pro1);
\draw [arrow] (pro1) -- (pro2);
\draw [arrow] (pro2) -- (dec1);
\draw [arrow] (dec1) -- node {no} (pro3a);
\draw [arrow] (dec1) -- node {yes} (pro3b);
\draw [arrow] (pro3a) -- (pro2);
\draw [arrow] (pro3b) -- (dec4);
\draw [arrow] (dec4) -- node {no} (pro1);
\draw [arrow] (dec4) -- node {yes} (stop);
\end{tikzpicture}
\caption{\label{flow_chart} Flow chart of the DMET algorithm.}
\end{figure}

The DMET algorithm with interacting bath is summarized in Fig. \ref{flow_chart}.

\section{Green's functions} \label{section_response_theory}
The DMET algorithm is not limited to ground-state properties, but can be extended to calculate response properties as well. In this section we review Ref.~\cite{georgePRB} in which the ground-state algorithm is extended to calculate Green's functions. In Ref.~\cite{georgePRB}, a bath space of many-body states was constructed by Schmidt decomposition of an approximate first order response. In this section, we will show that for the local density of states (LDOS), only one additional bath orbital is required. In other words, for the LDOS, the many-body bath space of Ref.~\cite{georgePRB} can be rewritten as a Fock space of bath orbitals.

Suppose one is interested in Green's functions of the form
\begin{equation}
 \mathcal{G}(\omega,\hat{X},\hat{V}) = \braket{ \Psi_0 \mid \hat{X}^{\dagger} \frac{1}{\omega - (\hat{H} - E_0) + i\eta} \hat{V} \mid \Psi_0 }.
\end{equation}
In the full Hilbert space, they can be calculated by first solving the linear problem
\begin{equation}
 \left( \omega - (\hat{H} - E_0) + i\eta \right) \ket{\Psi_1(\omega, \hat{V})} = \hat{V} \ket{ \Psi_0 },
\end{equation}
which subsequently yields $ \mathcal{G}(\omega,\hat{X},\hat{V}) = \braket{ \Psi_0 \mid \hat{X}^{\dagger} \mid \Psi_1(\omega, \hat{V})}$.
The same strategy is followed in Ref.~\cite{georgePRB}. First, an approximate bath space is constructed by Schmidt decomposition of
\begin{equation}
\ket{\Phi_1(\omega, \hat{V})} = \frac{1}{ \omega - ( \hat{h} + \hat{u} - \epsilon_{0} ) + i \eta } \hat{V} \ket{\Phi_0}, \label{dmet_response_fisrt_order}
\end{equation}
where $\epsilon_{0}$ is the ground-state energy of $\hat{h} + \hat{u}$ associated with the wavefunction $\ket{\Phi_0}$.
Subsequently, the linear problem is solved in the active space formed by the impurity and bath space:
\begin{equation}
 \left( \omega - (\hat{H^{\text{emb}}} - E^{\text{emb}}_0) + i\eta \right) \ket{\Psi^{\text{emb}}_1(\omega, \hat{V})} = \hat{V} \ket{ \Psi^{\text{emb}}_0 }. \label{dmet_response_fisrt_order2}
\end{equation}

In this formalism, $\hat{X}$ and $\hat{V}$ act on the impurity for which the embedded problem is formulated. The DMET correlation potential in Eqs.~\eqref{dmet_response_fisrt_order} and \eqref{dmet_response_fisrt_order2} is the ground-state one. As will be discussed in Sect.~\ref{sect_twodimhubbard}, accurate spectral functions are obtained with this method. However, in order to work with the many-body bath states arising in the Schmidt decomposition of Eq.~\eqref{dmet_response_fisrt_order}, the embedded Hamiltonian has to be constructed explicitly in the many-body basis. We will now show that for the LDOS, one can work with a Fock space of bath orbitals.

For LDOS, $\hat{X}$ and $\hat{V}$ are single-particle operators acting on a specific orbital. Here, we will only treat the addition part, but the discussion of the removal part is analogous. Eq.~\eqref{dmet_response_fisrt_order} becomes:
\begin{eqnarray}
\ket{\Phi_1(\omega, \hat{a}_k^{\dagger})}
    & = & \frac{1}{ \omega - ( \hat{h} + \hat{u} - \epsilon_{0} ) + i \eta } \hat{a}_k^{\dagger} \ket{\Phi_0}, \\
    & = & \sum\limits_{\mu}^{\text{virt}}\frac{1}{ \omega - ( \hat{h} + \hat{u} - \epsilon_{0} ) + i \eta } \hat{a}_{\mu}^{\dagger} C_{\mu k}^{\dagger} \ket{\Phi_0}, \\
    & = & \sum\limits_{\mu}^{\text{virt}}\frac{1}{ \omega - \epsilon_{\mu} + i \eta } \hat{a}_{\mu}^{\dagger} C_{\mu k}^{\dagger} \ket{\Phi_0},
\end{eqnarray}
where $\epsilon_{\mu}$ is the single-particle energy associated with virtual orbital $\mu$ of the single-particle operator $\hat{h} + \hat{u}$. We can now augment the ground-state bath orbital space with one additional orbital, arising from the added electron. This electron has weight on both the impurity and the environment:
\begin{equation}
\sum\limits_{\mu}^{\text{virt}} \frac{C_{\mu k}^{\dagger}}{ \omega - \epsilon_{\mu} + i \eta } \hat{a}_{\mu}^{\dagger} = \sum\limits_{l}^{L_{\text{A}}} \sum\limits_{\mu}^{\text{virt}} \frac{ C_{l \mu} C_{\mu k}^{\dagger}}{ \omega - \epsilon_{\mu} + i \eta } \hat{a}_{l}^{\dagger} + \sum\limits_{l > L_{\text{A}}} \sum\limits_{\mu}^{\text{virt}} \frac{ C_{l \mu} C_{\mu k}^{\dagger}}{ \omega - \epsilon_{\mu} + i \eta } \hat{a}_{l}^{\dagger}.
\end{equation}
The part on the environment can be added to the ground-state bath orbital space, after normalization similar to Eq.~\eqref{bathorbitalconstruction}. If real-valued instead of complex-valued orbitals are desired, two additional bath orbitals should be added, for the real and imaginary parts. The response \eqref{dmet_response_fisrt_order} is then spanned exactly in the Fock space of the augmented set of bath orbitals.

\section{Overview of the literature} \label{sect_applications}

Ground-state DMET has been applied to a variety of condensed matter systems. It has been used to study the one-dimensional Hubbard model \cite{geraldPRL, bulikPRB}, the one-dimensional Hubbard-Anderson model \cite{troyGeminals}, the one-dimensional Hubbard-Holstein model \cite{barbara}, the two-dimensional Hubbard model on the square \cite{geraldPRL, boxiaoHubbard, simons_collaboration} as well as the honeycomb lattice \cite{qiaoniPRB}, and the two-dimensional spin-$\frac{1}{2}$ $J_1$-$J_2$-model \cite{spinsystemPRB}.

Within the context of quantum chemistry, the method has been used to study hydrogen rings and sheets \cite{geraldJCTC, seb_dmet}, beryllium rings \cite{seb_dmet}, an S$_{\text{N}}$2 reaction \cite{seb_dmet}, polymers \cite{bulikJCP}, boron-nitride sheets \cite{bulikJCP}, and crystalline diamond \cite{bulikJCP}.

The DMET bath orbital construction from Sect.~\ref{bath_orbital_section} can also be used to construct optimal QM/MM boundaries \cite{qimingJCTC} and to contract primitive Gaussians into adaptive atomic basis sets for correlated calculations \cite{sorrellaJCP}.

The DMET formalism is not limited to ground-state properties. By augmenting the ground-state bath space with additional correlated many-body states from a Schmidt decomposition of the response wavefunction, accurate spectral functions have been obtained \cite{georgePRB, qiaoniPRB}. We have shown in Sect.~\ref{section_response_theory} that for the local density of states it is sufficient to augment the bath orbital space with one additional response orbital.

\section{The one-band Hubbard model on the square lattice} \label{sect_twodimhubbard}
In this section, we review the DMET calculations of Refs. \citenum{geraldPRL}, \citenum{boxiaoHubbard}, and \citenum{georgePRB} on the one-band Hubbard model on the square lattice. This model contains sufficient physics to exhibit $d$-wave superconductivity. For this reason, many groups have invested considerable numerical effort to map its rich phase diagram, which contains a Mott metal-insulator transition, $d$-wave superconductivity, and magnetism. For a detailed overview, we refer the reader to Refs. \citenum{scalapino} and \citenum{simons_collaboration}.

The Hubbard Hamiltonian
\begin{equation}
\hat{H}_{\text{Hubbard}} = -t \sum\limits_{<ij>,\sigma} \hat{a}^{\dagger}_{i \sigma} \hat{a}_{j \sigma}  -t' \sum\limits_{\ll ij \gg,\sigma} \hat{a}^{\dagger}_{i \sigma} \hat{a}_{j \sigma} + U \sum\limits_{i} \hat{n}_{i,\uparrow} \hat{n}_{i,\downarrow}
\end{equation}
is expressed in terms of spatial orbitals and contains nearest-neighbour ($t$) and next-nearest-neighbour ($t'$) hopping, as well as on-site electron repulsion ($U$).

\begin{figure}
 \centering
 \includegraphics[width=0.7\textwidth]{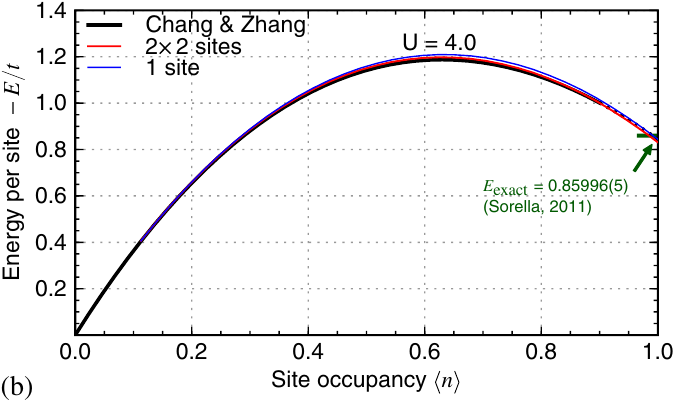}
 \caption{\label{gerald_prl_one} Reprinted figure with permission from G. Knizia and G. K.-L. Chan, \textit{Physical Review Letters} \textbf{109}, 186404 (2012) \cite{geraldPRL}. Copyright (2012) by the American Physical Society. Ground-state energy per site for the one-band Hubbard model on the square lattice with $U/t=4$ and nearest-neighbour hopping only ($t'=0$). DMET calculations with impurity sizes $1 \times 1$ and $2 \times 2$ are in very good agreement with earlier QMC results from Chang and Zhang \cite{PhysRevB.78.165101} and Sorella \cite{PhysRevB.84.241110}.}
\end{figure}

For half-filling ($n=1$) and nearest-neighbour hopping only ($t'=0$), the model has particle-hole symmetry. Quantum Monte Carlo (QMC) becomes exact in this regime, because there is no fermion sign problem. It is therefore instructive to compare DMET calculations with QMC. Fig.~\ref{gerald_prl_one} compares DMET results with QMC \cite{PhysRevB.78.165101, PhysRevB.84.241110} both at and away from half-filling. The results improve as the impurity size becomes larger, and local bulk properties such as the energy per site will eventually saturate with increasing impurity size. Tab.~\ref{boxiao_table} contains DMET results which are extrapolated with respect to impurity size, and compares them with AFQMC and DMRG \cite{simons_collaboration}. The DMET ground-state energies are in very good agreement with the QMC and DMRG results.

\begin{table}
\centering
 \begin{tabular}{lllll}
 \hline
 \hline
 $U/t$ & Filling & DMET & AFQMC & DMRG \\
 \hline
 2  & 1.0 & -1.1764(3)  & -1.1763(2) & -1.176(2)   \\
 4  & 1.0 & -0.8604(3)  & -0.8603(2) & -0.862(2)   \\
 6  & 1.0 & -0.6561(5)  & -0.6568(3) & -0.658(1)   \\
 8  & 1.0 & -0.5234(10) & -0.5247(2) & -0.5248(2)  \\
 12 & 1.0 & -0.3686(10) & -0.3693(2) & -0.3696(3)  \\
 4  & 0.8 & -1.108(2)   & -1.110(3)  & -1.1040(14) \\
 4  & 0.6 & -1.1846(5)  & -1.185(1)  &             \\
 4  & 0.3 & -0.8800(3)  & -0.879(1)  &             \\
 \hline
 \hline
 \end{tabular}
 \caption{\label{boxiao_table} Reprinted table with permission from B. Zheng and G. K.-L. Chan, \textit{Physical Review B} \textbf{93}, 035126 (2016) \cite{boxiaoHubbard}. Copyright (2016) by the American Physical Society. Ground-state energy per site for the one-band Hubbard model on the square lattice with nearest-neighbour hopping only ($t'=0$). The DMET results are extrapolated with respect to impurity size. For filling $n=1.0$ there is no fermion sign problem and the AFQMC results are exact. The DMET, AFQMC, and DMRG results \cite{simons_collaboration} are in very good agreement.}
\end{table}

\begin{figure}
 \centering
 \includegraphics[width=0.6\textwidth]{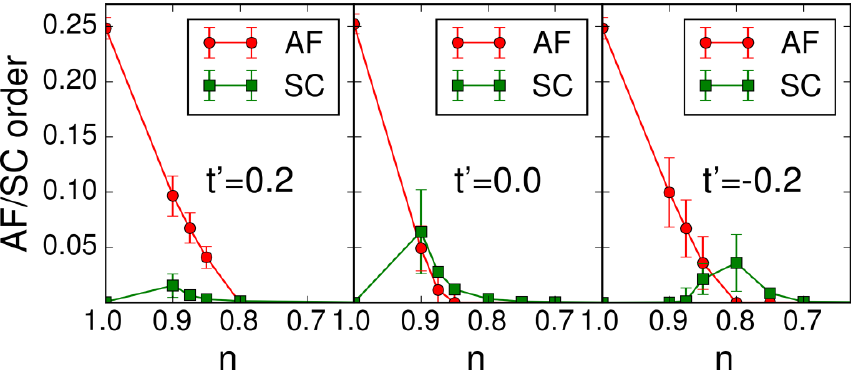}
 \caption{\label{boxiao_prb_one} Reprinted figure with permission from B. Zheng and G. K.-L. Chan, \textit{Physical Review B} \textbf{93}, 035126 (2016) \cite{boxiaoHubbard}. Copyright (2016) by the American Physical Society. DMET antiferromagnetic (AF) and superconducting (SC) order parameters for the one-band Hubbard model on the square lattice as a function of lattice filling $n$ for $U/t=4$. Near half-filling ($n=1.0$) antiferromagnetism is observed, and below half-filling $d$-wave superconductivity.}
\end{figure}

This gives confidence to study not only ground-state energies, but various properties as well. Fig.~\ref{boxiao_prb_one} shows the antiferromagnetic and superconducting order parameters as a function of lattice filling $n$ for $U/t=4$. Near half-filling antiferromagnetism is observed, and below half-filling $d$-wave superconductivity.

\begin{figure}
 \centering
 \includegraphics[width=0.6\textwidth]{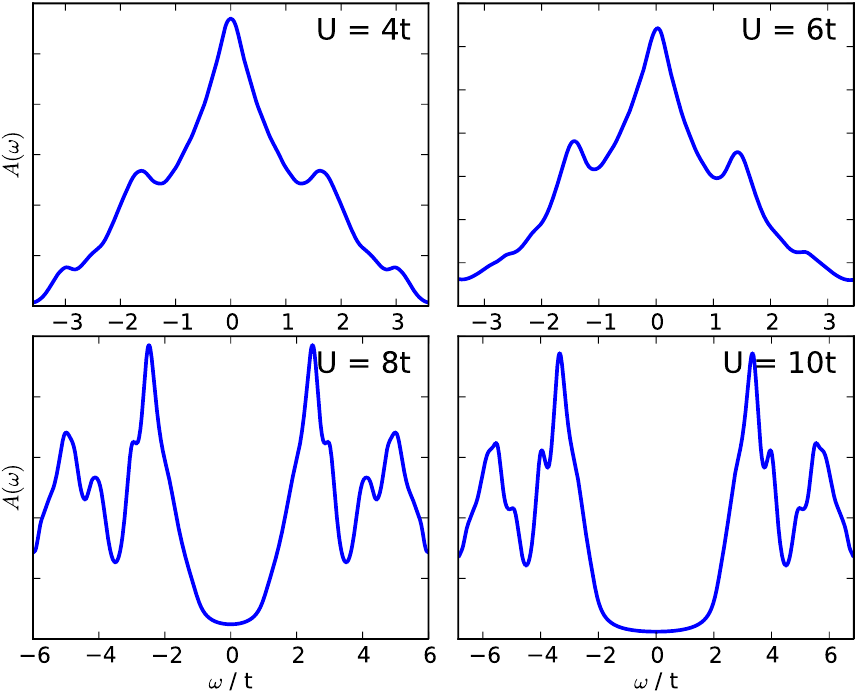}
 \caption{\label{george_prb_one} Reprinted figure with permission from G. H. Booth and G. K.-L. Chan, \textit{Physical Review B} \textbf{91}, 155107 (2015) \cite{georgePRB}. Copyright (2015) by the American Physical Society. Local density of states for the one-band Hubbard model on the square lattice at half-filling ($n=1$) and nearest-neighbour hopping only ($t'=0$). DMET Green's function calculations on an impurity of size $2 \times 2$ show the opening of a gap with increasing $U/t$, indicating a metal-insulator transition in the paramagnetic phase near $U/t \approx 6.9$.}
\end{figure}

The paramagnetic phase of the one-band Hubbard model on the square lattice can be studied by using \textit{restricted} Hartree-Fock theory as the low-level method. For this phase, the DMET correlation potential $\hat{u}$ opens a single-particle energy gap in $\hat{h} + \hat{u}$ with increasing $U/t$. This gap is a qualitative signature of the metal-insulator transition. With the Green's function DMET method explained in Sect.~\ref{section_response_theory} the local density of states can be calculated, which yields a quantitatively accurate gap. Fig.~\ref{george_prb_one} shows the opening of the gap with increasing $U/t$ for the one-band Hubbard model at half-filling ($n=1$) and nearest-neighbour hopping only ($t'=0$). A metal-insulator transition occurs near $U/t \approx 6.9$.

\section{Dissociation of a linear hydrogen chain} \label{sect_hydrogen_chaisn}

\begin{figure}
 \centering
 \includegraphics[width=0.75\textwidth]{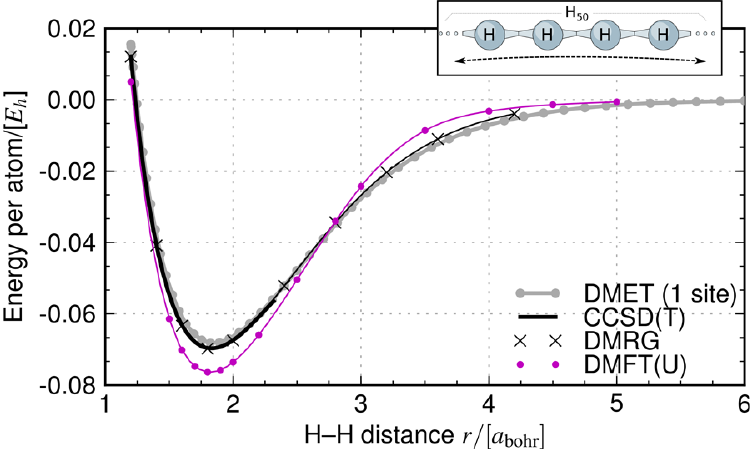}
 \caption{\label{gerald_jctc_one} Reprinted figure with permission from G. Knizia and G. K.-L. Chan, \textit{Journal of Chemical Theory and Computation} \textbf{9}, 1428-1432 (2013). Copyright (2013) by the American Chemical Society. Ground-state energy per atom for the symmetric stretch of a linear hydrogen chain with 50 atoms in the STO-6G basis. CCSD(T), DMFT \cite{PhysRevLett.106.096402}, and DMET calculations with one atom per impurity are compared with the numerically exact DMRG energies \cite{Hachmann}.}
\end{figure}

The dissociation of hydrogen chains has become a standard test case in quantum chemistry for strong correlation.
The ground-state energy per atom for the symmetric stretch of a linear hydrogen chain with 50 atoms in the STO-6G basis is shown in Fig.~\ref{gerald_jctc_one}. CCSD(T), DMFT \cite{PhysRevLett.106.096402}, and DMET calculations with one atom per impurity are compared with the numerically exact DMRG energies \cite{Hachmann}. DMET performs significantly better than DMFT for this case, most likely due to the interacting bath which represents long-range electron interactions between the impurity and the environment beyond mean-field, which are not present in the DMFT calculations.

\section{Summary}

The general goal of embedding theories is to obtain properties of interest of an impurity, without doing expensive calculations on the whole quantum system. When there is entanglement (strong correlation) between the impurity and the environment, the embedding method should be able to capture it. The most well-known embedding method for strong correlation is dynamical mean-field theory (DMFT), in which the correlated frequency-dependent Green's function of the impurity is determined self-consistently. DMFT is however computationally quite involved when only ground-state properties are of interest. A simpler and cheaper alternative is density matrix embedding theory (DMET), in which the impurity is embedded in an approximate many-body Schmidt basis for the environment, which is optimized self-consistently via the so-called DMET correlation potential.

In Sect.~\ref{entanglement_introduction} we have reviewed quantum entanglement and the Schmidt decomposition, two concepts from quantum information theory which are necessary to understand DMET. In Sect.~\ref{dmet_general_introduction} a general introduction to DMET is given. For mean-field low-level wavefunctions, the many-body Schmidt basis for the environment is a Fock space of bath orbitals. In Sect.~\ref{bath_orbital_section} we have outlined how these bath orbitals can be calculated. This leads to a DMET active space which consists of the impurity and bath orbitals. The embedding Hamiltonian for DMET can be constructed similar to the CASCI effective Hamiltonian, \textit{i.e.} the one-electron integrals contain Coulomb and exchange terms from the external core orbitals. In Sect.~\ref{embedded_hamiltonian} the construction of the embedded Hamiltonian and the calculation of DMET properties are discussed. The DMET correlation potential allows to fine-tune the bath orbital space and is optimized self-consistently as described in Sect.~\ref{section_self_consistency}. While DMET was introduced as a simpler and cheaper alternative to DMFT for ground-state properties, it is not limited to ground-state properties. In Sect.~\ref{section_response_theory} the extension of DMET to linear response and Green's functions is outlined. For the local density of states, the many-body Schmidt basis for the approximate response in the environment can be obtained by augmenting the ground-state bath orbital space with one additional response orbital. A general overview of the literature is given in Sect.~\ref{sect_applications}. DMET calculations on the one-band Hubbard model on the square lattice and the symmetric stretch of a hydrogen chain are reviewed in Sect.~\ref{sect_twodimhubbard} and \ref{sect_hydrogen_chaisn}, respectively.

DMET has shown success in describing condensed matter and quantum chemical systems where strong correlation is present. Although in principle, the method can also handle weak correlation, we have not discussed how well it can be handled with the outlined algorithm. We refer the reader to Ref.~\cite{seb_dmet} for a brief discussion.

\section*{Acknowledgements}
S. W. gratefully acknowledges a Gustave Bo\"el - Sofina - B.A.E.F. postdoctoral fellowship from the King Baudouin Foundation and the Belgian-American Educational Foundation for the academic year 2014-2015, and a postdoctoral fellowship from the Research Foundation Flanders (Fonds Wetenschappelijk Onderzoek Vlaanderen) for the academic years 2015-2018. G. K.-L. C. acknowledges support from the US Department of Energy through DE-SC0010530. Additional support was provided from the Simons Foundation through the Simons Collaboration on the Many-Electron Problem.


\bibliography{thebiblio}

\begin{thebibliography}{10}

\bibitem{geraldPRL}
Gerald Knizia and Garnet Kin-Lic Chan.
\newblock Density matrix embedding: A simple alternative to dynamical
  mean-field theory.
\newblock {\em Phys. Rev. Lett.}, 109(18):186404, 2012.

\bibitem{geraldJCTC}
Gerald Knizia and Garnet Kin-Lic Chan.
\newblock Density matrix embedding: A strong-coupling quantum embedding theory.
\newblock {\em J. Chem. Theory Comput.}, 9(3):1428--1432, 2013.

\bibitem{boxiaoHubbard}
Bo-Xiao Zheng and Garnet Kin-Lic Chan.
\newblock Ground-state phase diagram of the square lattice {H}ubbard model from
  density matrix embedding theory.
\newblock {\em Phys. Rev. B}, 93(3):035126, 2016.

\bibitem{simons_collaboration}
J.~P.~F. LeBlanc, Andrey~E. Antipov, Federico Becca, Ireneusz~W. Bulik, Garnet
  Kin-Lic Chan, Chia-Min Chung, Youjin Deng, Michel Ferrero, Thomas~M.
  Henderson, Carlos~A. Jim\'enez-Hoyos, E.~Kozik, Xuan-Wen Liu, Andrew~J.
  Millis, N.~V. Prokof'ev, Mingpu Qin, Gustavo~E. Scuseria, Hao Shi, B.~V.
  Svistunov, Luca~F. Tocchio, I.~S. Tupitsyn, Steven~R. White, Shiwei Zhang,
  Bo-Xiao Zheng, Zhenyue Zhu, and Emanuel Gull.
\newblock Solutions of the two-dimensional {H}ubbard model: Benchmarks and
  results from a wide range of numerical algorithms.
\newblock {\em Phys. Rev. X}, 5(4):041041, 2015.

\bibitem{troyGeminals}
Takashi Tsuchimochi, Matthew Welborn, and Troy Van~Voorhis.
\newblock Density matrix embedding in an antisymmetrized geminal power bath.
\newblock {\em J. Chem. Phys.}, 143(2):024107, 2015.

\bibitem{barbara}
B.~Sandhoefer and G.~K.-L. Chan.
\newblock Density matrix embedding theory for interacting electron-phonon
  systems.
\newblock {\em arXiv 1602.04195}, 2016.

\bibitem{spinsystemPRB}
Zhuo Fan and Quan-Lin Jie.
\newblock Cluster density matrix embedding theory for quantum spin systems.
\newblock {\em Phys. Rev. B}, 91(19):195118, 2015.

\bibitem{PhysRevLett.62.324}
Walter Metzner and Dieter Vollhardt.
\newblock Correlated lattice fermions in $d=\ensuremath{\infty}$ dimensions.
\newblock {\em Phys. Rev. Lett.}, 62(3):324--327, 1989.

\bibitem{PhysRevLett.69.1240}
Antoine Georges and Werner Krauth.
\newblock Numerical solution of the \textit{d} = \ensuremath{\infty} {H}ubbard
  model: {E}vidence for a {M}ott transition.
\newblock {\em Phys. Rev. Lett.}, 69(8):1240--1243, 1992.

\bibitem{RevModPhys.68.13}
Antoine Georges, Gabriel Kotliar, Werner Krauth, and Marcelo~J. Rozenberg.
\newblock Dynamical mean-field theory of strongly correlated fermion systems
  and the limit of infinite dimensions.
\newblock {\em Rev. Mod. Phys.}, 68(1):13--125, 1996.

\bibitem{zgidDMFT}
Dominika Zgid and Garnet Kin-Lic Chan.
\newblock Dynamical mean-field theory from a quantum chemical perspective.
\newblock {\em J. Chem. Phys.}, 134(9):094115, 2011.

\bibitem{PhysRev.43.830}
J.~K.~L. MacDonald.
\newblock Successive approximations by the {R}ayleigh-{R}itz variation method.
\newblock {\em Phys. Rev.}, 43(10):830--833, 1933.

\bibitem{seb_dmet}
S.~Wouters, C.~A. Jim\'enez-Hoyos, Q.~Sun, and G.~K.-L. Chan.
\newblock A practical guide to density matrix embedding theory in quantum
  chemistry.
\newblock {\em arXiv 1603.08443}, 2016.

\bibitem{iao_gerald2}
Gerald Knizia.
\newblock Intrinsic atomic orbitals: An unbiased bridge between quantum theory
  and chemical concepts.
\newblock {\em J. Chem. Theory Comput.}, 9(11):4834--4843, 2013.

\bibitem{qimingJCTC}
Qiming Sun and Garnet Kin-Lic Chan.
\newblock Exact and optimal quantum mechanics/molecular mechanics boundaries.
\newblock {\em J. Chem. Theory Comput.}, 10(9):3784--3790, 2014.

\bibitem{qiaoniPRB}
Qiaoni Chen, George~H. Booth, Sandeep Sharma, Gerald Knizia, and Garnet Kin-Lic
  Chan.
\newblock Intermediate and spin-liquid phase of the half-filled honeycomb
  {H}ubbard model.
\newblock {\em Phys. Rev. B}, 89(16):165134, 2014.

\bibitem{bulikJCP}
Ireneusz~W. Bulik, Weibing Chen, and Gustavo~E. Scuseria.
\newblock Electron correlation in solids via density embedding theory.
\newblock {\em J. Chem. Phys.}, 141(5):054113, 2014.

\bibitem{georgePRB}
George~H. Booth and Garnet Kin-Lic Chan.
\newblock Spectral functions of strongly correlated extended systems via an
  exact quantum embedding.
\newblock {\em Phys. Rev. B}, 91(15):155107, 2015.

\bibitem{bulikPRB}
Ireneusz~W. Bulik, Gustavo~E. Scuseria, and Jorge Dukelsky.
\newblock Density matrix embedding from broken symmetry lattice mean fields.
\newblock {\em Phys. Rev. B}, 89(3):035140, 2014.

\bibitem{sorrellaJCP}
S.~Sorella, N.~Devaux, M.~Dagrada, G.~Mazzola, and M.~Casula.
\newblock Geminal embedding scheme for optimal atomic basis set construction in
  correlated calculations.
\newblock {\em J. Chem. Phys.}, 143(24):244112, 2015.

\bibitem{scalapino}
D.~J. Scalapino.
\newblock {Numerical Studies of the {2D} {H}ubbard Model}.
\newblock In J.~R. Schrieffer and J.~S. Brooks, editors, {\em Handbook of
  High-Temperature Superconductivity: Theory and Experiment}, chapter~13, pages
  495--526. Springer, New-York, 2007.

\bibitem{PhysRevB.78.165101}
C.-C. Chang and S.~Zhang.
\newblock Spatially inhomogeneous phase in the two-dimensional repulsive
  hubbard model.
\newblock {\em Phys. Rev. B}, 78(16):165101, 2008.

\bibitem{PhysRevB.84.241110}
S.~Sorella.
\newblock Linearized auxiliary fields monte carlo technique: Efficient sampling
  of the fermion sign.
\newblock {\em Phys. Rev. B}, 84(24):241110, 2011.

\bibitem{PhysRevLett.106.096402}
N.~Lin, C.~A. Marianetti, A.~J. Millis, and D.~R. Reichman.
\newblock {Dynamical Mean-Field Theory for Quantum Chemistry}.
\newblock {\em Phys. Rev. Lett.}, 106:096402, 2011.

\bibitem{Hachmann}
J.~Hachmann, W.~Cardoen, and G.~K.-L. Chan.
\newblock Multireference correlation in long molecules with the quadratic
  scaling density matrix renormalization group.
\newblock {\em J. Chem. Phys.}, 125(14):144101, 2006.

\end{thebibliography}

\end{document}